\pdfoutput=1 
\documentclass[ 
 aps,%
 twocolumn,%
  reprint,%
 prl,%
 groupedaddress,%
 superscriptaddress,%
  showpacs,%
 lengthcheck,%
 amsfonts,%
 floatfix,%
]{revtex4-1} 

\RequirePackage{amsmath,amssymb,bm}
\RequirePackage{graphicx}
\RequirePackage[]{subfigure}
\RequirePackage{hyperref}
\hypersetup{%
  breaklinks = {true},
  citecolor = {blue},
  colorlinks = {true},
  linkcolor = {red},
  pdfauthor = {\textcopyright\ F\'{a}bio Hip\'{o}lito },
  pdfcreator = {\LaTeX\ and \flqq hyperref\frqq},
  pdffitwindow = {true},
  pdfmenubar = {true},
  pdfpagelayout = {SinglePage},
  pdfstartview = {Fit},
  pdftoolbar = {true},
  plainpages = {false},
}

\newcommand{\eg}{\emph{e.g.}}

\newcommand{\teq}{{\,=\,}}
\newcommand{\tneq}{{\,\neq\,}}
\newcommand{\tequiv}{{\,\equiv\,}}
\newcommand{\tsim}{{\,\sim\,}}
\newcommand{\tplus}{{\,+\,}}
\newcommand{\tminus}{{\,-\,}}

\newcommand{\tto}{{\,\to\,}}
\newcommand{\szero}{\ensuremath{\sigma_0}}

\usepackage[usenames,dvipsnames]{xcolor}

\RequirePackage[normalem]{ulem} 

\graphicspath{ {./Figs/} } 

\begin{document}

\title[SH spectroscopy to optically detect valley polarization in 2D materials]
{Second harmonic spectroscopy to optically detect valley polarization in 2D 
materials}

\author{F. Hipolito}

\affiliation{Department of Physics and Nanotechnology,
Aalborg University, DK-9220 Aalborg {\O}st, Denmark}

\affiliation{%
Centre for Advanced 2D Materials, National University of Singapore, 
6 Science Drive 2, Singapore 117546%
} 

\author{Vitor M. Pereira} 
\thanks{Corresponding author: vpereira@nus.edu.sg} 

 \affiliation{%
Centre for Advanced 2D Materials, National University of Singapore, 
6 Science Drive 2, Singapore 117546%
} 
\affiliation{%
Department of Physics, National University of Singapore,%
2 Science Drive 3, Singapore 117542%
}

\begin{abstract}
Valley polarization (VP), an induced imbalance in the populations of a 
multi-valley electronic system, allows emission of second harmonic (SH) 
light even in centrosymmetric crystals such as graphene.
Whereas in systems such as MoS$\mathrm{_2}$ or BN this adds to their intrinsic 
quadratic response, SH generation in a multi-valley inversion-symmetric crystal 
can provide a direct measure of valley polarization.
By computing the nonlinear response and characterizing theoretically the 
respective SH as a function of polarization, temperature, electron density, and 
degree of VP, we demonstrate the possibility of disentangling and individually 
quantifying the intrinsic and valley contributions to the SH.
A specific experimental setup is proposed to obtain direct quantitative 
information about the degree of VP and allow its remote mapping. This approach 
could prove useful for direct, contactless, real-space monitoring of valley 
injection and other applications of valley transport and valleytronics.
\end{abstract}




\maketitle

\section{Introduction}
Interactions of light with matter beyond linear response are a rich source of 
fundamentally interesting phenomena as well as of many-fold opportunities for 
applications \cite{Jackson1999, Born1986, Shen2002, Boyd2008}. 
In particular, the use of non-linear optical spectroscopy for characterizing 
the electronic properties of crystalline materials has emerged as a fruitful, 
simple and important technique because, among other advantages, it allows fast, 
non-invasive probing of electronic systems and is sensitive to intermediate 
coherent electronic transitions \cite{Shen2002, Boyd2008}. Compared to linear 
optical absorption, for example, in non-linear optical spectroscopy there is a 
larger freedom in utilizing the expanded set of selection rules and conditions 
involving the polarization dependence or polarization state, in order to 
extract more microscopic details with the same type of measurement 
\cite{Ivchenko2005, McIver2012}. 
We can understand this in the simplest and most general way by recalling that 
the $n$--th order response is governed by a $(n+1)$--rank tensor and that, for 
a given crystalline symmetry, the number of independent optical constants 
increases with the order \cite{Haussuhl2007}. Therefore, since a single 
frequency measurement at higher order of response can capture a larger number 
of independent quantities of the system, it more strongly constrains its 
microscopic details (e.g. the modeling of its bandstructure) while, at the 
same time, becomes a more versatile approach that is capable of probing a richer 
set of phenomena. This has a large potential for characterization and 
applications and, consequently, is of high interest.

Two-dimensional crystals such as graphene, boron nitride, and transition-metal 
dichalcogenides (TMD) have been shown to have a particularly strong non-linear 
optical response, especially given their atomically thin character. In the 
latter, second harmonic generation (SHG) is particularly robust 
\cite{Ivchenko2005, McIver2012, Wu2012} and is routinely used for simple 
characterization tasks such as identifying crystal orientation, uniformity, or 
layer number \cite{Shen1989, Dean2009, Dean2010a, McIver2012b, Malard2013, 
Kumar2013a, Li2013a, Hsu2014, Zhou2015}. With a setup that allows 
for translation of the beam along the sample, it becomes possible to 
spatially map the SHG by probing the sample in scanning mode with resolution 
limited by the spot size \cite{Malard2013, Kumar2013a, Zhou2015}.
In graphene, on the other hand, SHG is forbidden in equilibrium by its $D_{6h}$ 
point group symmetry (PGS). As discussed below, the vanishing quadratic response 
in graphene arises at the microscopic level from the exact cancellation of 
finite contributions from the $\mathbf{K}$ and $\mathbf{K}'$ valleys. 

Previous theoretical calculations, indicate that disrupting the valley 
cancellation by population imbalance can lead to a finite SH with estimated 
magnitudes on par with conventional nonlinear crystals \cite{Golub2014, 
Wehling2015}, which means that VP is expected to generate a very strong 
non-linear signal.
Golub and Tarasenko \cite{Golub2014} calculated the frequency dependent SH 
susceptibility from an explicit integration of the time-dependent density 
matrix using an effective Dirac description of the electronic structure of 
graphene; trigonal warping is explicitly included both in the effective 
Hamiltonian and in the coupling to light.
Wheling \emph{et al.} \cite{Wehling2015} computed the SH optical conductivity 
from a diagrammatic expansion of the current response to second order, thus 
expressing all quantities in terms of Green's functions. Their description is 
based on a tight-binding (TB) formulation of the electronic problem in 
graphene, including the TB derivation of the generalized coupling to light and 
the velocity operator in higher orders from a Peierls substitution in the 
hoppings.
The different methods and approximations used in these references lead to not 
entirely compatible results. Moreover the scope of these calculations is 
limited by other strict approximations, \eg\ zero temperature, very small VP
and, above all, pertain to (gapless) graphene only. 

Therefore, it becomes important to address the more general and rich problem of 
valley-induced SHG in 2D materials beyond graphene using an approach capable 
of addressing a more general set of conditions, such as finite temperature, 
variable carrier density, and polarization of the incoming and outgoing 
radiation. From the technical point of view, we address these characteristics 
(see \S\ref{sec:freq_dep}) within the length gauge formalism \cite{Aversa1995, 
Hipolito2016}.
To be encompassing and allow a controlled breaking of inversion symmetry, we 
study the quadratic optical response to light of a generic two-band electronic 
system on a honeycomb lattice. This allows us to quantify the interplay between 
intrinsic contributions and those induced by a finite VP as a function of 
frequency and polarization. At the qualitative level this choice allows us to 
interpolate between the behavior of graphene (gapless) and that of 
semiconducting TMD (gapped).

Since much effort is currently invested to theoretically and experimentally 
develop methods and concepts to harness the valley degree of freedom in
these and related systems for valleytronic applications \cite{Rycerz2007, 
Zeng2012, Xiao2012, Golub2011, Mak2012, Jiang2013a, Mak2014a, Yu2014, 
Jiang2014, Lee2015a}, it is important to establish practical, versatile and 
reliable probes able to quantify and track the degree of valley polarization 
(VP), just as, in spintronics, it is crucial to have probes capable of 
quantifying spin polarization, injection, relaxation, etc.

Light is a demonstrably effective means of inducing a VP in these materials 
\cite{Mak2012,Zeng2012,Cao2012} and, here, we discuss a specific proposal 
of its utilization as an effective qualitative and quantitative probe as well 
through SH spectroscopy. Even though SHG in otherwise SH-dark graphene provides 
direct access to the degree of an induced VP, our discussion extends from this 
case to systems with intrinsic SHG where the two effects can be present and 
contribute to independent second-order optical constants.
To be encompassing and allow a controlled breaking of inversion symmetry, we do 
that by studying the quadratic optical response to light of a generic two-band 
electronic system on a honeycomb lattice. This allows us to quantify the 
interplay between intrinsic contributions and those induced by a finite VP as a 
function of frequency and polarization.
As the key underlying physics is not dependent on specific microscopic details 
other than the crystal symmetry, we begin by discussing a specific experimental 
procedure that should allow one to use SHG as a useful probe in valleytronics 
and, subsequently, analyze the microscopic details of SHG in the framework 
discussed above.

\section{Fingerprint of valley polarization in SHG}
\begin{figure}
\centering
\includegraphics[width=1.0\columnwidth]{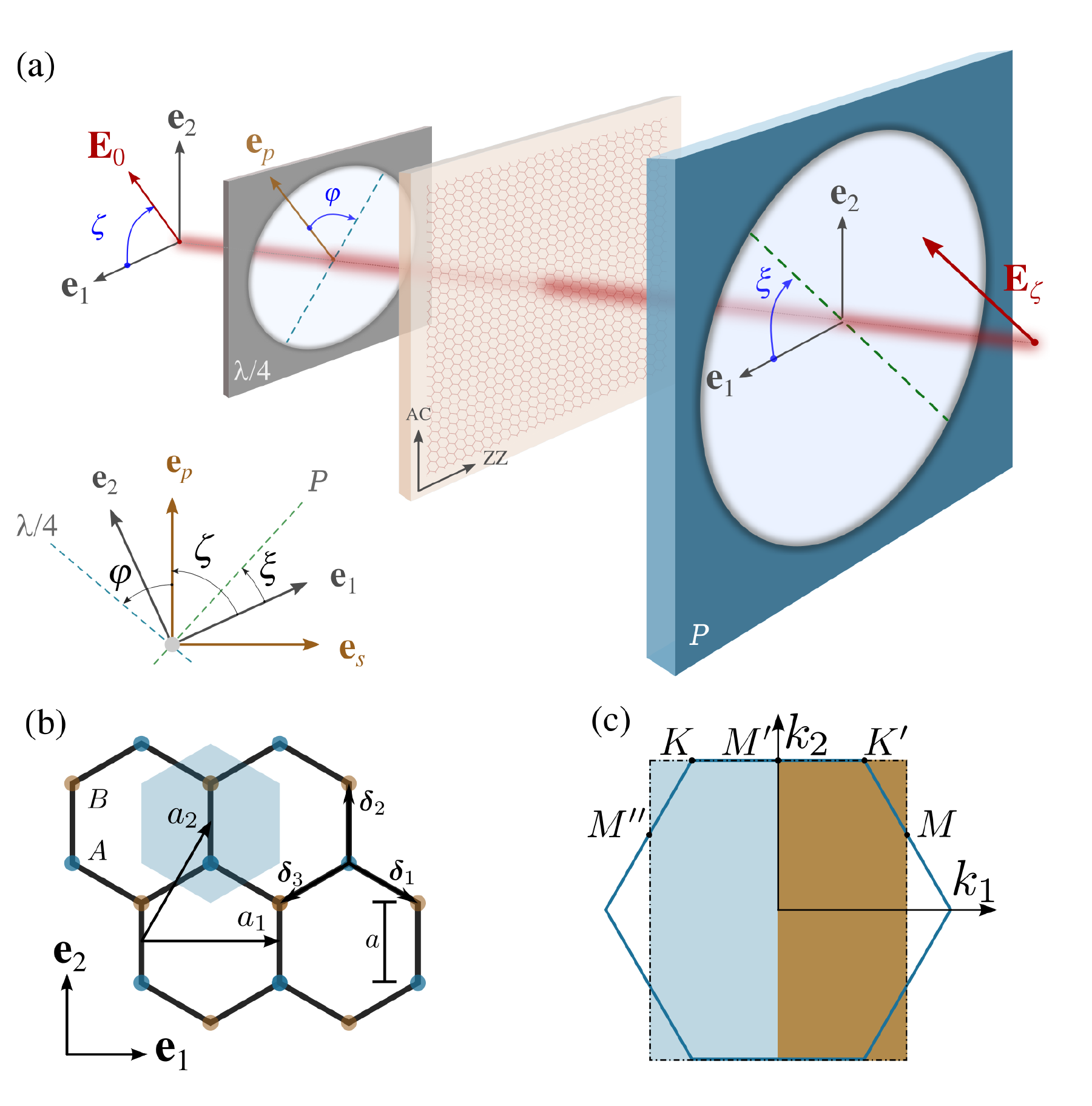}
\caption{(color online)
(a) Schematic setup for SH spectroscopy where light initially polarized 
linearly, $\textbf{E}_0$, passes through a $\lambda/4$ plate that sets the 
polarization state before impinging on the sample. The radiated SH is analyzed 
with a linear polarizer ($P$) before reaching the detector as 
$\mathbf{E}_\xi$. The bottom left shows the different angles discussed in the 
text ($\zeta$: plane of linear incoming polarization, $\varphi$: orientation of 
the $\lambda/4$ plate with respect to the incoming polarization, $\xi$: 
orientation of the output analyzer).
(b) Crystalline lattice and the choice of coordinate axes. 
(c) Brillouin zone where the light and dark rectangles represent its partition 
into areas associated with the \textbf{K} and \textbf{K}' valleys.
}
\label{fig:setup}
\end{figure}

Threefold rotational symmetry severely restricts the in-plane components of the 
quadratic conductivity that obey 
$\sigma_{\alpha\beta\beta} \teq \sigma_{\beta\alpha\beta} \teq 
\sigma_{\beta\beta\alpha} \teq \tminus\sigma_{\alpha\alpha\alpha}$, where 
$\alpha,\beta {\,\in\,} \lbrace 1, 2 \rbrace$ \cite{Haussuhl2007}. Alone, this 
symmetry reduces the number of independent components to just $\sigma_{111}$ and 
$\sigma_{222}$ which, for simplicity, we shall replace by the dimensionless 
counterparts $\bar\sigma_i \tequiv \sigma_{iii}/\szero$, where $\szero \tequiv 
e^3a/4\gamma_0\hbar$ sets the natural scale of the second order 2D conductivity 
(see below) \cite{Hipolito2016}. Furthermore, in a honeycomb lattice in 
equilibrium whose mirror plane $\mathbf{e}_m$ is parallel to $\mathbf{e}_2$ 
[fig.~\ref{fig:setup}(b)], only $\bar\sigma_2$ survives which defines the 
\emph{intrinsic} quadratic response of the system.
Since at frequencies much smaller that the bandwidth  
($\omega{\,\ll\,}\gamma_0$) electronic processes are governed by states in the 
vicinity of the two inequivalent points $\mathbf{K}$ and $\mathbf{K}'$ 
in the Brillouin zone (BZ), that intrinsic response is the combination of the 
contributions from each of these two valleys, which contribute independently 
(additively) in a translationally-invariant system.
A \emph{crucial} aspect, though, is that the PGS of $\mathbf{K}/\mathbf{K}'$ is 
still $D_{3h}$ \emph{but} with the mirror plane \emph{perpendicular} to that of 
the real space lattice ($\mathbf{e}_m$). In other words, if taken 
independently, 
each valley contributes $\bar\sigma_{1}{\,\neq\,}0$, $\bar\sigma_{2}{\,=\,}0$ 
and it is \emph{their sum} that yields an overall $\bar\sigma_{1}{\,=\,}0$, 
$\bar\sigma_{2}{\,\neq\,}0$ expected on symmetry grounds at equilibrium (in 
particular, if inversion symmetry is furthermore present as in graphene, the 
two valleys exactly cancel each other and 
$\bar\sigma_{1}{\,=\,}\bar\sigma_{2}{\,=\,}0$) \cite{Golub2011}. 
However, if there exists an imbalance in the population of the valleys, there 
is no symmetry constraint to quench either $\bar\sigma_{1}$ or 
$\bar\sigma_{2}$; in particular, SHG can arise through $\bar\sigma_{1}\tneq 0$ 
in a lattice with inversion symmetry, which immediately suggests the detection 
of this valley-induced SHG response as a direct optical probe of VP. 
Generically, in a VP crystal without inversion both components will be present 
with a direct impact on the polarization dependence of the SHG that we explore 
here to disentangle and independently quantify the intrinsic ($\bar\sigma_2$) 
and valley-induced ($\bar\sigma_1$) SHG.
We first derive the dependence of the SHG in the polarization state of the 
excitation field in general terms to establish the procedure for the individual 
component extraction, and afterwards analyze the frequency dependence of 
$\bar\sigma_{1}$ and $\bar\sigma_{2}$ in a microscopic model for graphene and 
for gapped graphene that applies qualitatively to the response in TMD.

We begin with the generic parameterization of an incoming monochromatic field 
$E_0$ normal to the sample, as illustrated in fig.~\ref{fig:setup}(a), that is 
initially \textit{p}-polarized before transmitting through a $\lambda/4$ plate 
with fast axis at an angle $\varphi$ with the plane of propagation. This 
permits 
the selection of any incoming polarization state, including linear 
polarization. 
For a general orientation of the propagation plane ($\zeta$) the (complex) 
amplitude of the electric field reaching the sample, $\mathbf{E}_{\omega}$, 
reads
$ \mathbf{E}_\omega \teq E_0\, \big( a \sin\zeta +b \cos\zeta
, \, -a \cos\zeta +b \sin\zeta , \, 0 \big) $
where $ a \tequiv i\sin( 2\varphi )/\sqrt{2}$, $ b \tequiv [ 1 \tminus i\cos( 
2\varphi ) ]/\sqrt{2} $ and $\mathbf{e}_{1}$ is aligned with the lattice zigzag 
direction. The second order two-dimensional current density, $ j_i^{(2)} ( 
\omega_1, \omega_2 ) = \sum_{jk} \sigma_{ijk}^{(2)} E_{\omega_1}^j 
E_{\omega_2}^k $, can hence be written as
\begin{subequations}
\begin{align}
\label{eq:jPheno}
j_1^{(2)} ( \omega_1, \omega_2 )  &= \szero
( f_1 \bar\sigma_1 +f_2 \bar\sigma_2 )
E_{\omega_1} E_{\omega_2} /2,
\\
j_2^{(2)} ( \omega_1, \omega_2 ) &= \szero
( f_2 \bar\sigma_1 -f_1 \bar\sigma_2 )
E_{\omega_1} E_{\omega_2} /2,
\end{align}
\end{subequations}
where the auxiliary functions $f_1$ and $f_2$ read
\begin{subequations}
\begin{align}
f_1 &\equiv 2\sin( 2\zeta + 2\varphi ) \sin( 2\varphi ) -2i\cos( 2\zeta 
+2\varphi ),
\\
f_2 &\equiv 2\cos( 2\zeta +2\varphi ) \sin( 2\varphi ) +2i\sin( 2\zeta 
+2\varphi ).
\end{align}
\end{subequations}
Even though we will be focusing on SHG arising from a single monochromatic 
source ($\omega_1\teq\omega_2\teq\omega$), we explicitly distinguish $\omega_1$ 
and $\omega_2$ to underline that our analysis applies to any second-order 
process. The sheet current eq.~\ref{eq:jPheno} radiates, in turn, an 
electromagnetic field with a flux density 
$I=\mu_0c\,|j^{(2)}(\omega_1,\omega_2)|^2/8$, or 
\begin{align}
I/ I_0 &=
 \big( |f_1|^2 +| f_2 |^2 \big)
 \big( | \bar\sigma_1 |^2  +| \bar\sigma_2 |^2 \big)
\nonumber \\ &-
8i ( \bar\sigma_1 \bar\sigma_2^*-\bar\sigma_1^* \bar\sigma_2 )
\sin( 2\varphi ),
\end{align}
where $I_0 \teq \mu_0c\, \szero^2 \;
|E_{\omega_1}|^2|E_{\omega_2}|^2/32 \teq (\mu_0c)^3 \szero^2
I_{\omega_1} I_{\omega_2}/8$ $[\mathrm{W/m^2}]$.
Whereas this shows that the total SHG intensity cannot discriminate the 
relative magnitudes of $\bar\sigma_1$ and $\bar\sigma_2$, that can be 
achieved by filtering the SH field with a linear polarizer parallel to the 
sample and rotated by $\xi$ with respect to $\mathbf{e}_1$ so that the electric 
field at the detector reads $\mathbf{E}_\xi \teq \mathbf{E}_{2\omega} 
{\,\cdot\,} (\cos\xi, \sin\xi, 0)$.
If the incoming light is linearly polarized parallel to the analizer 
($\xi\teq\zeta$) the SH intensity at the detector reads
\begin{align}
\label{eq:I:zeta}
I_{\parallel}/I_0 &=
  4|\bar\sigma_1|^2 \big[ \! \sin^2( 3\zeta \! +\! 2\varphi)\sin^2(2\varphi)  
  \! +\! \cos^2( 3\zeta \! +\! 2\varphi ) \big]
\nonumber \\ &
+ 4|\bar\sigma_2|^2 \big[ \! \cos^2( 3\zeta \! +\! 2\varphi)\sin^2(2\varphi)  
  \! +\! \sin^2( 3\zeta \! +\! 2\varphi ) \big]
\nonumber \\ &
-2( \bar\sigma_1 \bar\sigma_2^* +\bar\sigma_1^* \bar\sigma_2 ) 
\sin(6\zeta +4\varphi ) \cos^2( 2\varphi )
\nonumber \\ &
-4i( \bar\sigma_1 \bar\sigma_2^* -\bar\sigma_1^* \bar\sigma_2 ) 
\sin(2\varphi )
,
\end{align}
while at cross orientation ($\xi \teq \zeta \tplus 90^\circ$) it is given by 
eq.~\ref{eq:I:zeta} with the replacement $ 3\zeta \tto 3\zeta +90^\circ $.

Eq.~\ref{eq:I:zeta} or any of its variants can thus be used to directly 
obtain $\sigma_1$ and $\sigma_2$ as well as the orientation of the lattice by 
fitting experimental SH intensities as a function of polarization. This is a 
concept similar to the usage of SHG as a remote, non-invasive probe of lattice 
orientation, layer number and other properties in recent applications of 
two-dimensional crystals having intrinsic SHG \cite{Shen1989, Dean2009, 
Dean2010a, McIver2012b, Malard2013, Kumar2013a, Li2013a, Kumar2013a, 
Hsu2014, Zhou2015}. What we now propose and explicitly show is that, in 
addition, it follows from the general features of eq.~\ref{eq:I:zeta} that the 
same concept can be applied to monitor and quantify the presence of VP, which 
is 
of clear interest for applications envisaged in the realm of valleytronics, 
somewhat similarly to the uses of the Kerr rotation to monitor and map spin 
accumulation in spintronics \cite{Kato2004}.

\begin{figure*}
\centering
\subfigure[]{%
  \includegraphics[width=.4\linewidth]{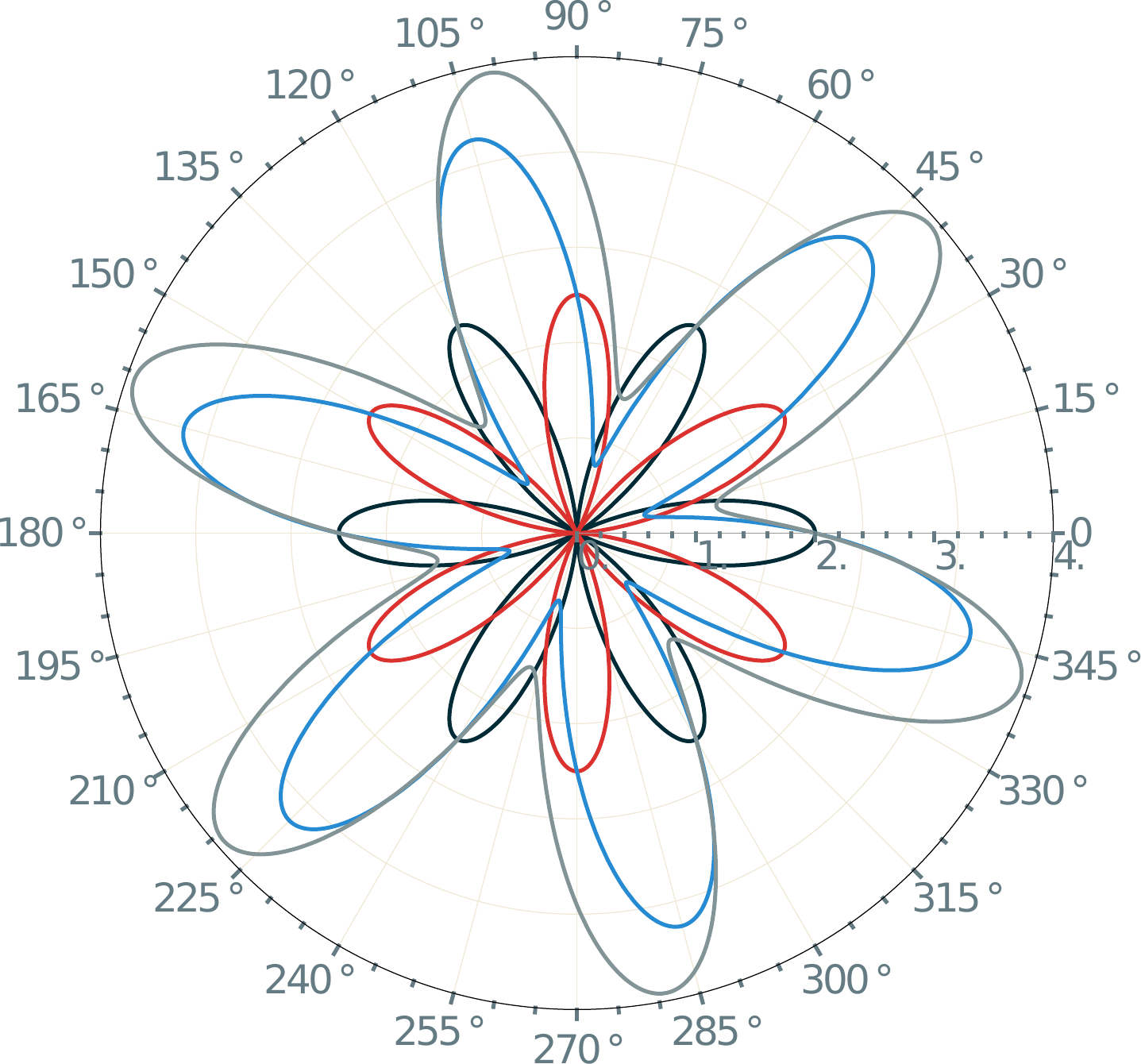}
  \label{fig:3fold:a}}
\hspace{.1\linewidth}
\subfigure[]{%
  \includegraphics[width=.4\linewidth]{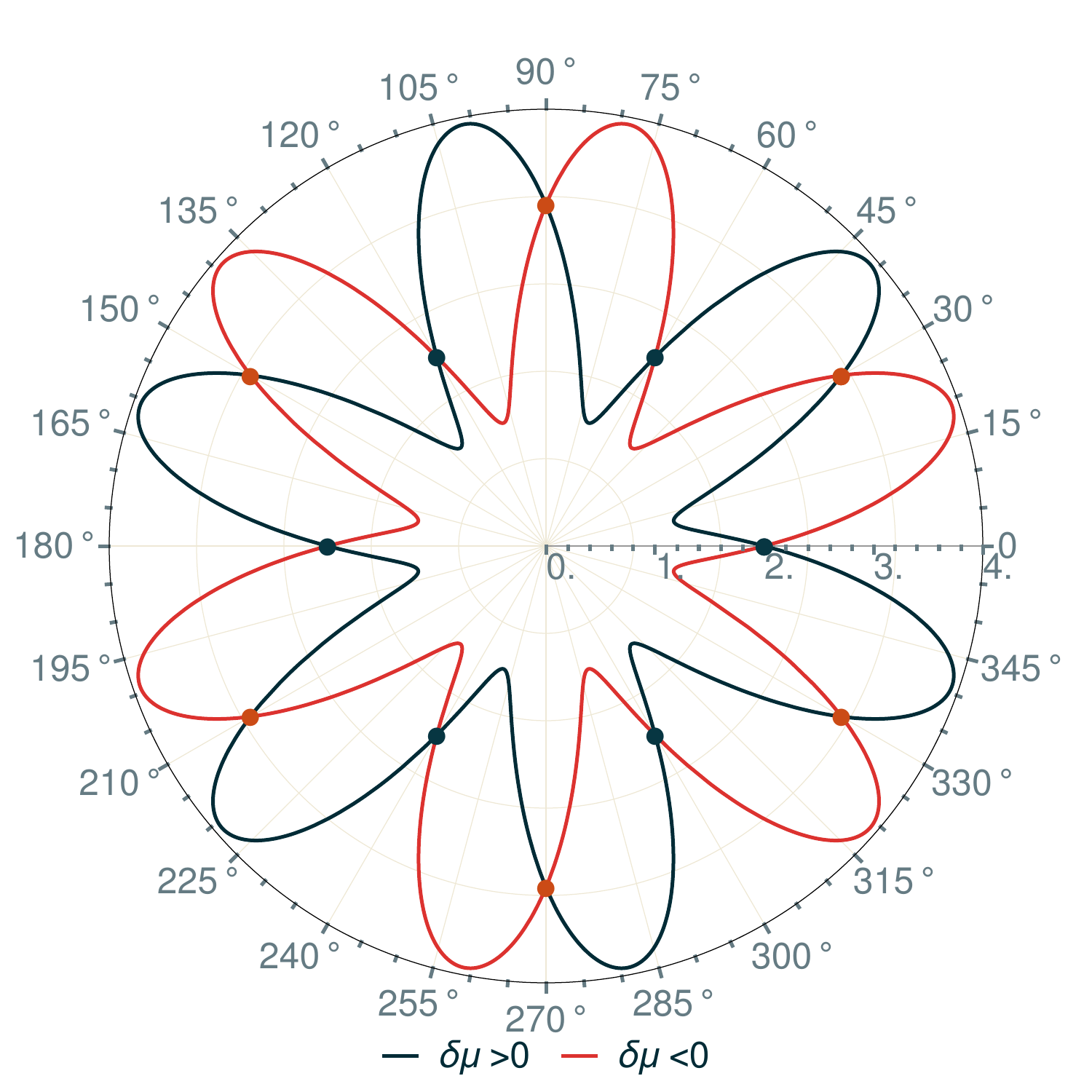}
  \label{fig:3fold:b}}
\caption{\label{fig:3fold}
Normalized SH intensity parallel to the polarization plane,
 $I_\parallel / I_0$, as a function of the polarization angle with respect to 
$\mathbf{e}_1$.
We consider $\varphi\teq 0$ so that the incoming light reaches the sample 
linearly polarized.
In (a), each curve represents a different scenario: 
$\bar\sigma_1 \teq 0$ (red),
$\bar\sigma_2 \teq 0 $ (black),
$\bar\sigma_2 / \bar\sigma_1 \teq e^{-i\pi/4} $ (blue),
$\bar\sigma_2 \teq \bar\sigma_1 e^{-i\pi/3} 5/4$ (gray).
In (b) we illustrate the reflection of the pattern about $\mathbf{e}_1$ under a 
reversal of the valley polarization ($\delta\mu {\,\to\,}\tminus\delta\mu$) 
with the same parameters used in the gray curve of (a).
}
\end{figure*}%

Fig.~\ref{fig:3fold:a} illustrates the two simplest and extreme cases of 
entirely intrinsic (red) and entirely VP-induced SHG (black).
For definiteness, we consider $\varphi\teq 0$ so that the incoming light 
reaches the sample linearly polarized.
As exactly anticipated from the earlier discussion on the differences between 
the PGS of the crystal and that of each valley, the angular pattern of 
$I_\parallel$ is rotated by 30$^{\circ}$ between the two cases: VP in an 
inversion symmetric crystal ($\bar\sigma_2\teq 0$) leads to SHG whose intensity 
is a direct measure of the degree of polarization $\mu_\mathbf{K} - 
\mu_{\mathbf{K}'}$ (more details below), and its flower-shaped pattern directly 
reveals the PGS of the $\mathbf{K}$ points.
The more general case of a system already having an intrinsic SHG 
($\bar\sigma_2\tneq 0$) is illustrated by the blue and gray curves. They reveal 
that an emerging VP is signaled by three distinct features:
(i) the progressive rotation of the flower pattern away from the principal
directions set by the lattice orientation; 
(ii) the increase in intensity as the contributions arising from $\bar\sigma_1$ 
add to the intrinsic SH intensity, as per eq.~\ref{eq:I:zeta}; 
(iii) the minimal intensity is no longer zero. Since the zero of the intrinsic 
SH response is usually well resolved experimentally \cite{Hsu2014},
any of these effects can be used for qualitative monitoring of the degree of VP 
in the system, a fast alternative to fitting the angular patterns to 
eq.~\ref{eq:I:zeta} when the actual magnitudes of $\bar\sigma_{1,2}$ are not 
required.
Note that $\bar\sigma_{1,2}$ are complex quantities and, hence, the orientation 
and shape of the pattern is determined not just by their relative magnitudes 
but also the relative phase, as easily seen in the case 
$\bar\sigma_1\teq\bar\sigma_2e^{i\delta}$: $ I_\parallel/I_0 \teq 
2|\bar\sigma_1|^2 [ 1 +\cos\delta\,\sin(6\zeta) ] $
(notably, when $\delta=\pm90^\circ$ the six-fold pattern vanishes and becomes 
isotropic). 

An obvious but important implication of these modifications is that the 
orientation of the SH intensity pattern \emph{does not} correlate directly 
anymore with the lattice orientation in the presence of \emph{both} intrinsic 
and VP-induced SHG. But this same fact can be utilized to detect and quantify 
both intrinsic and valley-induced conductivities.
As $\bar\sigma_1$ is to leading order linear in $\delta\mu$, a reversal of the 
VP ($\delta\mu\tto\tminus\delta\mu$) changes its sign. On account of the 
cross-term in eq.~\ref{eq:I:zeta}, this translates into a rotation of the 
pattern 
by 30$^{\circ}$, which is equivalent to a reflection about the principal 
directions set by the lattice orientation, as shown in fig.~\ref{fig:3fold:b}. 
Consequently, the intersection of two patterns associated with opposite VP 
defines the orientation of the lattice modulo 30$^{\circ}$. While this still 
doesn't uniquely distinguish ZZ and AC directions, we note that a unique 
identification is possible whenever $|\bar\sigma_1|\ne|\bar\sigma_2|$ (which 
comprises essentially all cases) because the two non-equivalent intersections 
will then occur at different SH intensities, and it follows from 
eq.~\ref{eq:I:zeta} that the intersection at lower (higher) intensity, 
highlighted with black (red) markers in the plot, occurs along the direction 
$\mathbf{e}_1{\,\Leftrightarrow\,}$ZZ ($\mathbf{e}_2{\,\Leftrightarrow\,}$AC) 
when $|\sigma_2|{\,>\,}|\sigma_1|$. Conversely, when 
$|\sigma_2|{\,<\,}|\sigma_1|$ the lower (higher) intersection occurs along 
$\mathbf{e}_2{\,\Leftrightarrow\,}$AC ($\mathbf{e}_1{\,\Leftrightarrow\,}$ZZ).
This is clearly seen in fig.~\ref{fig:3fold:b} where the reversal of 
$\delta\mu$ 
allows the immediate conclusion that the direction $\mathbf{e}_1$ corresponds 
to $\zeta=0$ because the two curves intersect there with the lowest intensity.
These considerations are relevant not just because they illustrate how to use 
all the available information for a facile and expedite characterization of 
the nonlinear optical constants, but also because the success of a full 
nonlinear fit of an experimental trace of $I_\parallel$ vs $\zeta$ to 
eq.~\ref{eq:I:zeta} can depend strongly on the assumed alignment of the lattice.
Finally, it is clear from eq.~\ref{eq:I:zeta} that, if the lattice orientation 
is 
known, measuring $I_\parallel/I_0$ at three non-equivalent orientations such as 
$\zeta\teq0^\circ,15^\circ,30^\circ$ suffices to uniquely determine the 
magnitudes of $\sigma_1$ and $\sigma_2$, as well as their relative phase.

The discussion so far was done for a linearly polarized excitation field 
($\varphi=0$). An alternative consists in analyzing the SH signal as a 
function of the polarization state of the excitation field 
determined by $\varphi$, and which can be tuned continuously with the rotation 
of a $\lambda/4$ plate \cite{Ganichev2003, Ivchenko2005, McIver2012}. Since the 
roles of $\varphi$ and $\zeta$ are very much equivalent in eq.~\ref{eq:I:zeta}, 
an analysis analogous to the one above can be straightforwardly done in this 
case. For example, with a fixed analyzer at $\xi\teq\xi_{zz}\tequiv 0$ 
($\xi_{ac} \tequiv 30^\circ$), the intensity can still be read from 
eq.~\ref{eq:I:zeta} with the replacement $3\zeta \tto 2\zeta$ ($3\zeta \tto 
2\zeta 
+90^\circ$). Since the description of the $\varphi$-dependence is similar to 
the one above, we omit it for brevity.

\section{Frequency dependence of \texorpdfstring{$\bar\sigma_1$}{sigma\_1} and 
\texorpdfstring{$\bar\sigma_2$}{sigma\_2}}
\label{sec:freq_dep}
In order to determine the typical dependence of both $\bar\sigma_1$ and 
$\bar\sigma_2$ on excitation frequency and chemical potential for 
representative 
cases, we focus the analysis now on graphene-based systems, where 
recent reports have demonstrated the possibility of generating valley-polarized 
currents with high valley relaxation lengths, both in mono and bilayers 
\cite{Gorbachev2014,Shimazaki2015}. The electronic degrees of freedom of a 
graphene monolayer are extremely well described by a single-orbital 
tight-binding (TB) model for electrons in the honeycomb lattice of 
fig.~\ref{fig:setup}(b). This is a single (hopping) parameter model for 
graphene, 
which can only have $\bar\sigma_1{\,\ne\,} 0$. 
In addition, in order to study the characteristics and relative magnitudes of 
$\bar\sigma_1$ and $\bar\sigma_2$ in a non inversion-symmetric system, it is 
desirable to have a model where inversion symmetry can be broken in a 
controlled 
way. That is easily incorporated in the single-orbital tight-binding via a 
sublattice potential $\pm\Delta/2$ which explicitly breaks the sublattice 
symmetry, and allows one to study the effects of VP in a more general ``gapped 
graphene'' setting. 
Whereas the case of graphene should be captured with good accuracy within this 
framework, the case of ``gapped graphene'' is expected to convey the main 
qualitative features expected in gapped systems such as in doped MoS$_2$ 
(doping suppresses excitonic effects, and renders a single-particle 
description of the optical response appropriate). 
The second order conductivity tensor is computed perturbatively for a 
translationally-invariant system treating the interaction with light via the 
direct coupling, $\mathbf{r \cdot E}$, in the dipole approximation as described 
in references \cite{Aversa1995,Hipolito2016}. We consider only the clean 
limit, but account phenomenologically for disorder broadening of the 
conductivity. 
Each component $\bar\sigma^{(2)}_{\lambda\alpha\beta}(\omega_1,\omega_2)$ is 
obtained from the formal result (25) of reference \cite{Hipolito2016}. 

\begin{figure}[t]
\centering
\includegraphics[width=\columnwidth]{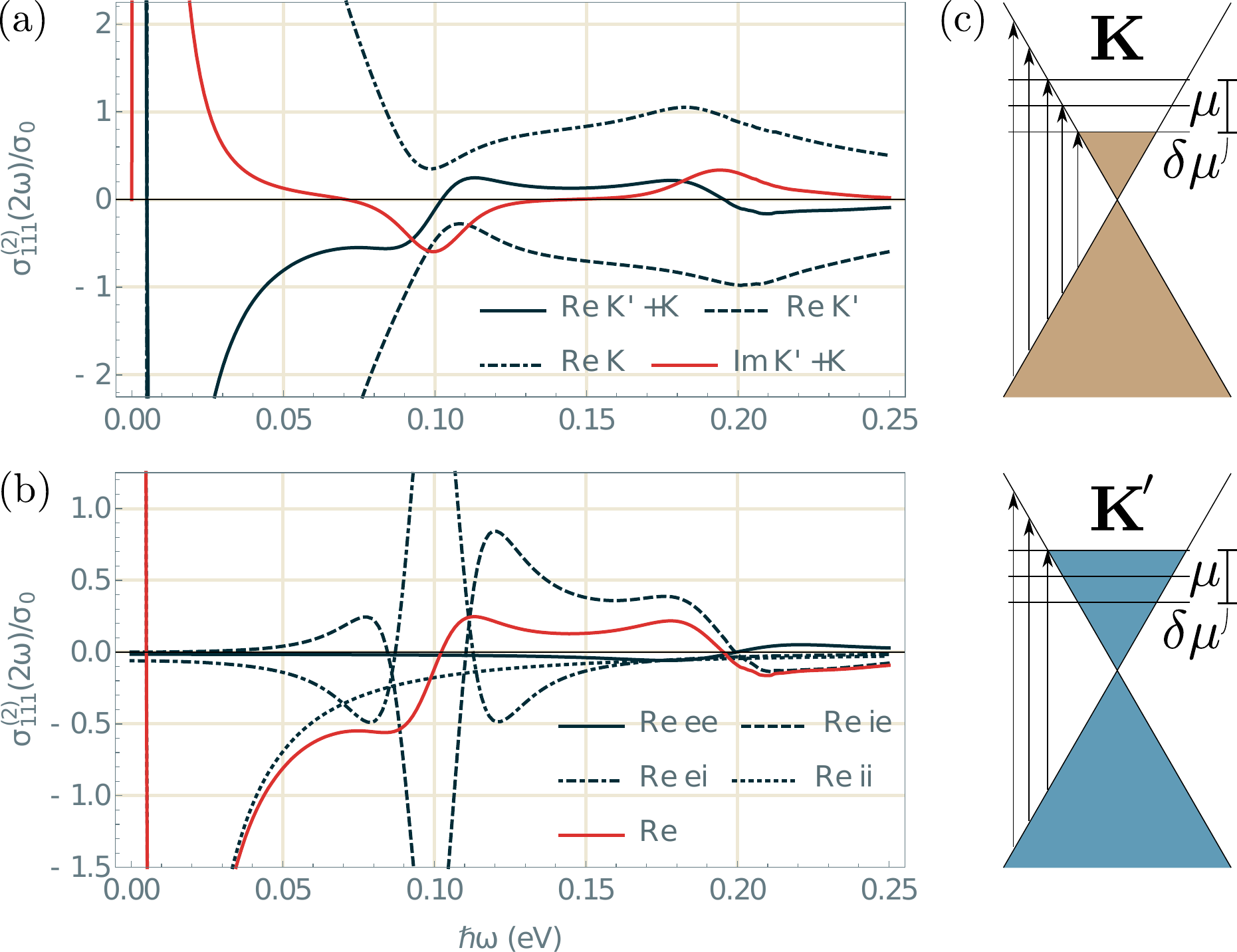}
\caption{(color online)
Second order conductivity in valley-polarized graphene.
The solid lines in (a) show the real and imaginary parts, while dashed and 
dot-dashed lines are the independent contributions from each valley.
(b) shows how the total response decomposes in terms of the various 
inter/intra-band processes.
$[ \gamma_0\teq 3$\,eV, $\Gamma\teq 5$\,meV, $T\teq 50$\,K, $\mu\teq 
0.1$\,eV, $\delta\mu\teq 10$\,meV, $\sigma_0\teq 2.88 \times 10^{-15} \, 
\mathrm{Sm/V}]$.
In (c) we depict the allowed transitions in each valley at $T \teq 0$\,K.
}
\label{fig:VP:0}
\end{figure}%

Our results show explicitly that, as expected, the $\mathbf{k}$-space 
integration for small photon energies $\hbar\omega \leq \gamma_0$, is dominated 
by the vicinity of the \textbf{K} points.
This allows us to use the equilibrium results to compute the contribution of 
each separate valley at different chemical potential by restricting the 
momentum integration to either of the shaded regions in fig. 
\ref{fig:setup}(c), 
while still working with the full tight-binding. 
Being able to keep the full tight-binding bandstructure rather than a 
Dirac-type approximation is important because, on the one hand, this allows us 
to immediately accommodate any refinement of the bandstructure model or 
straightforwardly extend the analysis to a different material. On the other 
hand, SHG in the clean limit arises only when the trigonal warping of the bands 
is explicitly considered \cite{Golub2014}, which is guaranteed in the TB scheme.

\begin{figure}[t]
\centering
\includegraphics[width=\columnwidth]{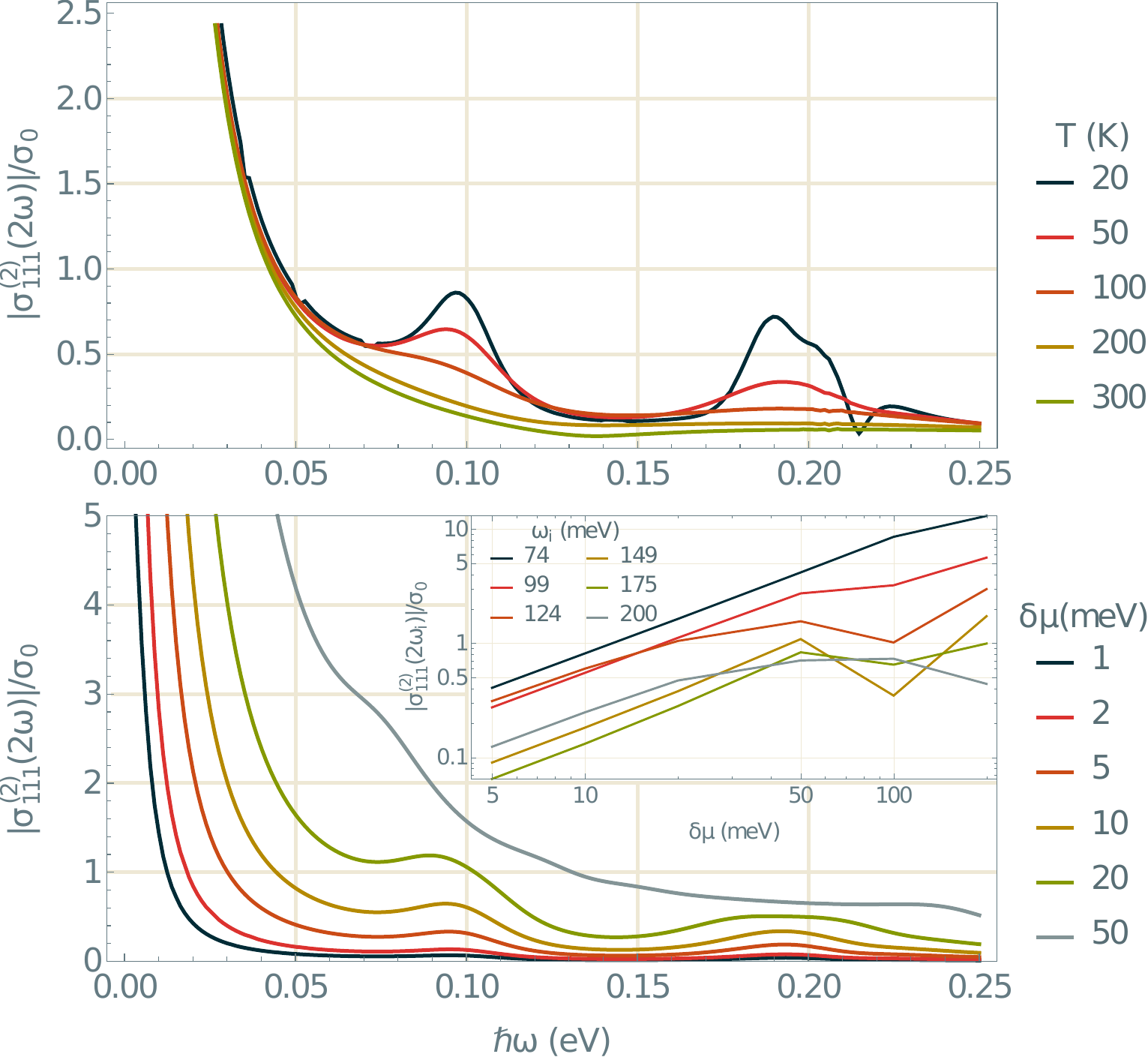}
\caption{(color online)
The effect of temperature and degree of VP on the second order conductivity of 
graphene.
The top panel has $\sigma_1$ at different temperatures with fixed $\delta\mu 
\teq 10$\,meV, showing how the features in the vicinity of 
$\omega\tsim\mu,\,2\mu$ are strongly temperature-dependent.
The second panel shows the dependence on $\delta\mu$ at $T\teq 50$\,K.
$ [ \gamma_0\teq 3$\,eV, $\Gamma\teq 5$\,meV, $\mu\teq 100$\,meV$ ]$
}
\label{fig:VP:1}
\end{figure}%

Starting with the case of graphene, it is instructive to consider first the 
individual contribution of each valley.
Since the point group symmetry of \textbf{K} has no inversion, each valley 
contributes a finite SHG (through $\bar\sigma_1\tneq 0$) but, in equilibrium 
($\mu_\mathbf{K}=\mu_{\mathbf{K}'}$), time-reversal symmetry forces an exact 
cancellation of each valley's contribution and a system such as graphene has no 
intrinsic SHG (cf. dashed curves in the top panel of fig.~\ref{fig:VP:0}). 
When a VP is induced as in fig.~\ref{fig:VP:0}, there is no cancellation anymore
and the overall effect at finite frequency is the appearance of two features at 
$\omega\teq\mu$ and $2\mu$ and a strong enhancement when $\omega\to0$, 
consistent with two previous reports based on a related calculation in the 
Dirac approximation \cite{Golub2014,Wehling2015}. 
The bottom panel shows the decomposition of $\bar\sigma_1$ in terms of the 
inter and intraband contributions defined in references 
\cite{Aversa1995,Hipolito2016}
\footnote{We follow the notation of \cite{Aversa1995,Hipolito2016}, where 
$ee$ represents purely interband, $ie$ and $ei$ mixed inter-intraband 
processes, and $ii$ purely intraband processes. These results are in 
qualitative agreement in with \cite{Golub2014}, but exhibit magnitudes 
$\tsim500$-fold larger, as the intensity of the resonant features increases 
significantly at low temperature.}%
.
Whereas the behavior at $\hbar\omega\tsim\mu$ and $2\mu$ is due to the resonant 
denominators coming from interband processes, the signal is much amplified 
towards the DC limit because the lack of inversion within each valley implies 
that when $\omega\tto 0$ the response is dominated by purely intraband 
transitions, since the conductivity terms $\sigma_1^{\mathrm{(ii)}}$ are 
now strictly finite (they cancel by symmetry in equilibrium 
\cite{Hipolito2016}).
We note that, despite being at the same level of single-particle approximation, 
our results in fig.~\ref{fig:VP:0} disagree with the previous calculations of 
the 
SHG in valley-polarized graphene \cite{Golub2014, Wehling2015} (which, in turn, 
themselves disagree with each other \footnote{With respect to Ref. 
\cite{Wehling2015}, despite capturing the divergence 
${\propto\,}1/\omega^2$ in the DC limit, our results are qualitatively 
different 
in the range $\mu\lesssim\hbar\omega\lesssim 2\mu$ in what regards the shape of 
the resonances at $\mu$ and $2\mu$, as well as the relative sign of 
$\bar\sigma_1$ in the two limits $\omega\to\,0,\,\infty$. Regarding the results 
of Ref. \cite{Golub2014}, there is some qualitative agreement, but a 
detailed analysis shows that the results do not match quantitatively in the 
amplitude and shape of the resonances.}).
We attribute these differences to the long-standing problem of taking proper 
account of the intra-band contributions in the calculation of nonlinear 
response functions. Since a VP leads to explicitly finite intra-band terms even 
in the presence of a band-gap (see below), such contributions must be handled 
with care, which is addressed here in the framework of Aversa and Sipe that has 
been proven reliable in the DC limit \cite{Blount1962, Sipe1993, Aversa1995}. 

The dependence of this valley-induced SHG on $T$ and $\mu$ is addressed in 
figure fig.~\ref{fig:VP:1}. The features at $\omega\teq\mu$ and $2\mu$ are 
strongly 
temperature-dependent and disappear as soon as $k_B T{\,\gtrsim\,}\delta\mu$ 
(top panel) because, at this point, the temperature broadening of the 
Fermi-Dirac function whittles down the effective valley polarization.
At fixed $T$, the response grows linearly with $\delta\mu$ \cite{Golub2014} 
when $\omega{\,\lesssim\,}\mu$, except near the resonances at $\omega 
\tsim \mu$ and $2\mu$.
%

\begin{figure}[t]
\centering
\includegraphics[width=\columnwidth]{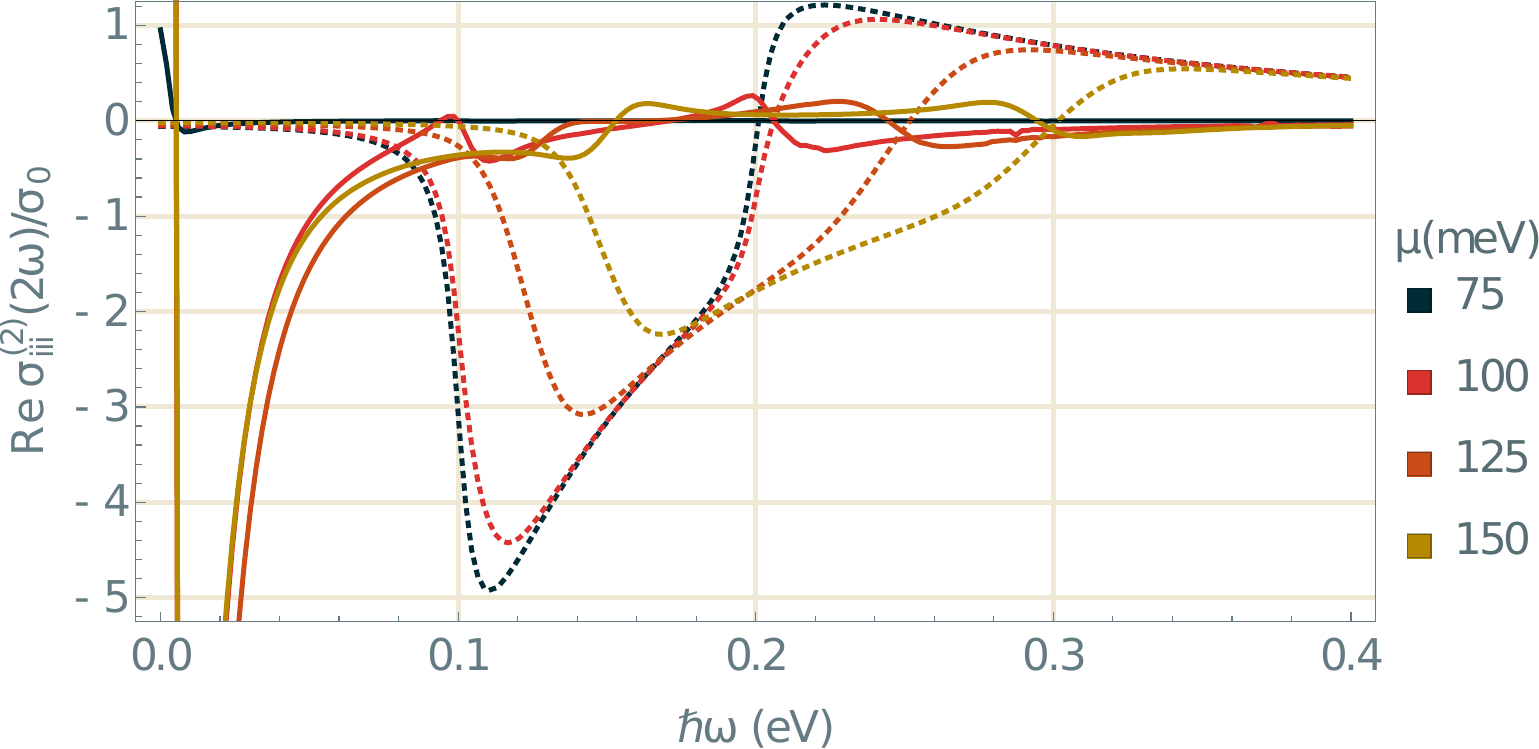}
\caption{(color online)
Second order conductivity in gapped graphene with VP at finite doping.
Solid (dashed) lines represent $\bar\sigma_1$ ($\bar\sigma_2$) at different 
$\mu$.
$ [ \gamma_0 \teq 3 \, \mathrm{eV}$, $\Delta 
\teq 200 \, \mathrm{meV}$, $\Gamma \teq 5 \,\mathrm{meV}$, $T \teq 50
\,\mathrm{K}$, $\delta\mu \teq 10 \, \mathrm{meV} ]$
}
\label{fig:VPgap}
\end{figure}%

Setting $\Delta\tneq 0$ explicitly breaks the inversion symmetry of graphene 
and 
an intrinsic SHG ($\bar\sigma_2\tneq 0$) obtains. For definiteness, consider 
the case when $\Delta/2$ and $\mu$ are comparable, which we illustrate in 
fig.~\ref{fig:VPgap} for $\delta\mu \teq 10$\,meV at $T\teq 50$\,K.
We see that lower frequencies $\omega \leq \mu \tsim \Delta/2$ are dominated by 
the VP mechanism ($|\bar\sigma_1|{\,\gg\,}|\bar\sigma_2| $), while the 
intrinsic response dominates for $\omega{\,\gtrsim\,\mu}$. This happens because 
$\bar\sigma_2$ is Pauli-blocked at $\omega{\,\lesssim\,}\mu$ but, as seen 
above, 
$\bar\sigma_1$ is enhanced at lower frequencies and varies weakly with $\mu$ 
(except if $|\mu\pm\delta\mu|{\,\approx\,}\Delta/2$, cf. black curve, since 
then 
VP is strongly affected by small changes in $\mu$). As a result, even in a 
situation where $\bar\sigma_1$ and $\bar\sigma_2$ might have comparable 
maximum magnitudes, it is possible to separate the VP-dominated and 
intrinsic-dominated regimes by tuning the relative position of the excitation 
frequency and $\mu$, since the latter can be used to push up the spectral 
region for which $\mathrm{Re}\,\sigma_2\ne0$. 
Furthermore, the rapid whittling of $\bar\sigma_1$ when $k_B T{\,\gtrsim\,} 
\delta\mu$, in contrast with the robustness of $\bar\sigma_2$ up to 
temperatures, significantly above room temperature \cite{Hipolito2016} results 
in a strong temperature dependence of the ratio $|\bar\sigma_1| / 
|\bar\sigma_2|$.

\section{Concluding remarks}
We studied the generation and polarization dependence of SH in threefold 
symmetric 2D materials with a finite VP, and performed specific 
microscopic calculations of the SH conductivity for a model that applies 
(accurately) to graphene and (qualitatively) to semiconducting TMD such as 
MoS$_2$.
Our results show that VP and intrinsic (when present) quadratic response 
generate distinct contributions with contrasting symmetry properties, which can 
be disentangled by analyzing the dependence of SHG on the orientation of 
polarization plane [cf. eq.~\ref{eq:I:zeta}] or on the state of polarization 
$\varphi$.
To achieve this, the SHG signal $I_\parallel$ can be used to determine the 
orientation of the lattice by either reversing the sign of the VP, or by 
probing 
the dependence on $\zeta$ at photon energies above (below) $\mu$, in the regime 
dominated by intrinsic (VP) where the maxima indicate the armchair (zigzag) 
directions of the lattice.
Knowledge of the lattice orientation (thus obtained or otherwise), permits a
direct application of eq.~\ref{eq:I:zeta} to extract the two independent 
nonlinear optical constants $\bar\sigma_1$ and $\bar\sigma_2$. Since 
$\bar\sigma_1$ is proportional to $\delta\mu$, the SH fingerprint of these 
systems can be used to directly identify and quantify an underlying imbalance 
between the populations in the valleys $\mathbf{K}$ and $\mathbf{K}'$.

If performed with a small spot size in a scanning mode, such measurements 
provide a means to directly map valley polarization throughout a system, 
measure the spatial decay of valley currents, and investigate the possibility 
or efficacy of their injection across heterostructure junctions and interfaces.

Our data for $\sigma_{iii}$ is presented in units of $\sigma_0\teq 2.88 \times 
10^{-15}$\,S\,m/V. If converted to 3D quadratic susceptibilities using an 
effective graphene thickness of $d\tequiv3.4$\,\AA, this corresponds to 
$\chi^{(2)} \tequiv \sigma_0 /(\omega \varepsilon_0d) {\,\approx\,} 6.3 \, 
\mathrm{nm/V}$ at $\omega\teq 0.1$\,eV.
As a reference, $\chi^{(2)}$ in a good non-linear bulk crystal has typical 
values of $0.01 \, \mathrm{nm/V}$ (ZnO) \cite{Chan2006}, 
$0.5\,\mathrm{nm/V}$ (GaAs, MoS$_2$) \cite{Bergfeld2003,Nastos2006,Wagoner1998}, 
$2 \, \mathrm{nm/V}$ (monolayer GaSe) \cite{Zhou2015}.
Hence, SHG due to valley polarization can exceed largely the typical non-linear
response of bulk materials such as GaSe.
The ability to vary the reference Fermi level in most atomically thin 
crystals through gating \cite{Novoselov2005a}, when combined with frequency 
dependent measurements, further expands the versatility of SH spectroscopy to 
assess valley-dependent properties, rendering it a valuable characterization 
tool in the nascent field of valleytronics. 

\acknowledgments 

FH thanks A. H. Castro Neto and M. Milletar\`{i} for their support and 
discussions throughout this project. 
He was supported by the National Research Foundation Singapore under the 
CRP award NRF-CRP6-2010-05 and by the QUSCOPE center sponsored by the Villum 
Foundation. VMP was supported by the Singapore Ministry of Education through 
grant MOE2015-T2-2-059.
Numerical computations were carried out at the HPC facilities of the NUS Centre 
for Advanced 2D Materials.

\bigskip

\bibliography{valley-pol}

\end{document}